\documentclass[11pt]{article}
\usepackage{amssymb,amsmath,amsfonts}
\usepackage{graphicx}
\usepackage{graphics}
\usepackage{eepic,epsfig}

\begin{document}

\title{Fermionic Casimir effect in toroidally compactified \\ de Sitter
spacetime}
\author{A. A. Saharian\thanks{%
E-mail: saharian@ictp.it } \\
%EndAName
\textit{Department of Physics, Yerevan State University, }\\
\textit{1 Alex Manogian Street, 0025 Yerevan, Armenia }}
\maketitle

\begin{abstract}
We investigate the fermionic condensate and the vacuum expectation values of
the energy-momentum tensor for a massive spinor field in de Sitter spacetime
with spatial topology $\mathrm{R}^{p}\times (\mathrm{S}^{1})^{q}$. Both
cases of periodicity and antiperiodicity conditions along the compactified
dimensions are considered. By using the Abel-Plana formula, the topological
parts are explicitly extracted from the vacuum expectation values. In this
way the renormalization is reduced to the renormalization procedure in
uncompactified de Sitter spacetime. It is shown that in the uncompactified
subspace the equation of state for the topological part of the
energy-momentum tensor is of the cosmological constant type. Asymptotic
behavior of the topological parts in the expectation values is investigated
in the early and late stages of the cosmological expansion. In the limit
when the comoving length of a compactified dimension is much smaller than
the de Sitter curvature radius the topological part in the expectation value
of the energy-momentum tensor coincides with the corresponding quantity for
a massless field and is conformally related to the corresponding flat
spacetime result. In this limit the topological part dominates the
uncompactified de Sitter part. In the opposite limit, for a massive field
the asymptotic behavior of the topological parts is damping oscillatory for
both fermionic condensate and the energy-momentum tensor.
\end{abstract}

\bigskip

\section{Introduction}

In recent years much attention has been paid to the possibility that a
universe could have non-trivial topology \cite{Lach95,Levi02}. Many of high
energy theories of fundamental physics are formulated in higher dimensional
spacetimes and it is commonly assumed that the extra dimensions are
compactified. In particular, the idea of compactified dimensions has been
extensively used in supergravity and superstring theories. From an
inflationary point of view universes with compact spatial dimensions, under
certain conditions, should be considered a rule rather than an exception
\cite{Lind04}. The models of a compact universe with non-trivial topology
may play an important role by providing proper initial conditions for
inflation (for physical motivations of considering compact universes see
also \cite{Star98}). There has been a large activity to search for
signatures of non-trivial topology by identifying ghost images of galaxies,
clusters or quasars. Recent progress in observations of the cosmic microwave
background provides an alternative way to observe the topology of the
universe \cite{Levi02}. If the scale of periodicity is close to the particle
horizon scale then the changed appearance of the microwave background sky
pattern offers a sensitive probe of the topology.

The compactification of spatial dimensions leads to a number of interesting
quantum field theoretical effects which include instabilities in interacting
field theories \cite{Ford80a}, topological mass generation \cite%
{Ford79,Toms80a,Toms80b}, symmetry breaking \cite{Toms80b,Odin88}. In the
case of non-trivial topology the boundary conditions imposed on fields give
rise to the modification of the spectrum for vacuum fluctuations and, as a
result, to the Casimir-type contributions in the vacuum expectation values
of physical observables (for the topological Casimir effect and its role in
cosmology see \cite{Most97,Bord01} and references therein). Compactification
of extra dimensions have moduli parameters which parametrize the size and
the shape of the extra dimensions and the Casimir effect has been used to
stabilize these moduli. The Casimir energy can also serve as a model for
dark energy needed for the explanation of the present accelerated expansion
of the universe (see \cite{Milt03,Eliz06,Gree07} and references therein).

De Sitter (dS) spacetime is among the most important cosmological
backgrounds. There are several physical motivations for this. In most
inflationary models an approximately dS spacetime is employed to solve a
number of problems in standard cosmology \cite{Lind90}. More recently
astronomical observations of high redshift supernovae, galaxy clusters and
cosmic microwave background \cite{Ries07} indicate that at the present epoch
the universe is accelerating and can be well approximated by a world with a
positive cosmological constant. If the universe would accelerate
indefinitely, the standard cosmology would lead to an asymptotic dS
universe. In addition to the above, an interesting topic which has received
increasing attention is related to string-theoretical models of dS spacetime
and inflation. Recently a number of constructions of metastable dS vacua
within the framework of string theory are discussed (see, for instance, \cite%
{Kach03,Silv07} and references therein). dS spacetime is the maximally
symmetric solution of the Einstein equations with a positive cosmological
constant and due to its high symmetry numerous physical problems are exactly
solvable on this background. A better understanding of physical effects in
this background could serve as a handle to deal with more complicated
geometries.

As it was argued in Refs. \cite{McIn04}, there is no reason to believe that
the version of dS spacetime which may emerge from string theory, will
necessarily be the most familiar version with symmetry group $O(1,4)$ and
there are many different topological spaces which can accept the dS metric
locally. There are many reasons to expect that in string theory the most
natural topology for the universe is that of a flat compact three-manifold.
The quantum creation of the universe having toroidal spatial topology is
discussed in \cite{Zeld84} and in references \cite{Gonc85} within the
framework of various supergravity theories. The effect of the
compactification of a single spatial dimension in dS spacetime (topology $%
\mathrm{R}^{D-1}\times \mathrm{S}^{1}$) on the properties of quantum vacuum
for a scalar field with general curvature coupling parameter and with
periodicity condition along the compactified dimension is investigated in
Ref. \cite{Saha07} (for quantum effects in braneworld models with dS spaces
and in higher-dimensional brane models with compact internal spaces see, for
instance, Refs. \cite{dSbrane,Flac03}). More general classes of the
compactification with topology $\mathrm{R}^{p}\times (\mathrm{S}^{1})^{q}$
in the case of scalar fields with both periodicity and antiperiodicity
conditions are investigated in \cite{Bell08}. Note that for a scalar field
on the background of uncompactified dS spacetime the renormalized vacuum
expectation values of the field square and the energy-momentum tensor are
investigated in Refs. \cite{Cand75,Dowk76,Bunc78} (see also \cite{Birr82}).
The corresponding effects upon phase transitions in an expanding universe
are discussed in \cite{Vile82,Alle83}. Trace anomaly for higher spin fields
in dS spacetime is considered in Ref. \cite{Birr79}.

In the present paper, partly motivated by possible applications in
supersymmetric theories, we investigate one-loop quantum effects for a
fermionic field on background of 4-dimensional dS spacetime with spatial
topology $\mathrm{R}^{p}\times (\mathrm{S}^{1})^{q}$. Among the most
important quantities characterizing the properties of the fermionic vacuum
are the fermionic condensate and the expectation value of the
energy-momentum tensor. Though the corresponding operators are local, due to
the global nature of the vacuum, these quantities describe the global
properties of the bulk and carry an important information about the
topology. In addition, the vacuum expectation value of the energy-momentum
tensor acts as a source of gravity in the Einstein equations and, hence,
plays an important role in modelling self-consistent dynamics involving the
gravitational field. We have organized the paper as follows. In the next
section we consider the plane wave fermionic eigenfunctions for the problem
under consideration. In sections \ref{sec:FermCond} and \ref{sec:EMT} these
eigenfunctions are used for the evaluation of the fermionic condensate and
the vacuum expectation values of the energy-momentum tensor. The behavior of
these quantities is investigated in various asymptotic regions of the
parameters. In section \ref{sec:AntiPer}, a fermionic field with
antiperiodiciy conditions along the compactified dimensions (twisted field)
is considered. The last section contains a summary of the work.

\section{Plane wave eigenspinors in de Sitter spacetime with compact spatial
dimensions}

\label{sec:EigFunc}

The dynamics of a massive fermionic field in curved spacetime with the
metric tensor $g^{\mu \nu }$ is governed by the Dirac equation
\begin{equation}
i\gamma ^{\mu }\nabla _{\mu }\psi -m\psi =0\ ,  \label{Direq}
\end{equation}%
with the covariant derivative operator
\begin{equation}
\nabla _{\mu }=\partial _{\mu }+\Gamma _{\mu }\ .  \label{Covder}
\end{equation}%
Here $\gamma ^{\mu }=e_{(a)}^{\mu }\gamma ^{(a)}$ are the Dirac matrices in
curved spacetime and $\Gamma _{\mu }$ is the spin connection given in terms
of the flat space Dirac matrices $\gamma ^{(a)}$ by the relation
\begin{equation}
\Gamma _{\mu }=\frac{1}{4}\gamma ^{(a)}\gamma ^{(b)}e_{(a)}^{\nu }e_{(b)\nu
;\mu }\ .  \label{Gammamu}
\end{equation}%
Note that in this formula $;$ means the standard covariant derivative for
vector fields. In the equations above $e_{(a)}^{\mu }$ is the tetrad field
satisfying the relation $e_{(a)}^{\mu }e_{(b)}^{\nu }\eta ^{ab}=g^{\mu \nu }$%
, where $\eta ^{ab}$ is the Minkowski spacetime metric tensor. In the
discussion below the flat space Dirac matrices will be taken in the standard
form%
\begin{equation}
\gamma ^{(0)}=\left(
\begin{array}{cc}
1 & 0 \\
0 & -1%
\end{array}%
\right) ,\;\gamma ^{(a)}=\left(
\begin{array}{cc}
0 & \sigma _{a} \\
-\sigma _{a} & 0%
\end{array}%
\right) ,\;a=1,2,3,  \label{gam0l}
\end{equation}%
with $\sigma _{1},\sigma _{2},\sigma _{3}$ being the Pauli matrices.

We consider a quantum fermionic field on background of 4-dimensional de
Sitter spacetime described by the line element%
\begin{equation}
ds^{2}=dt^{2}-e^{2t/\alpha }\sum_{i=1}^{3}(dz^{i})^{2},  \label{ds2deSit}
\end{equation}%
and having spatial topology $\mathrm{R}^{p}\times (\mathrm{S}^{1})^{q}$, $%
p+q=3$. The Ricci scalar and the corresponding cosmological constant are
related to the parameter $\alpha $ in the expression for the scale factor by
the formulae%
\begin{equation}
R=12/\alpha ^{2},\;\Lambda =3/\alpha ^{2}.  \label{Ricci}
\end{equation}%
For the further discussion, in addition to the synchronous time coordinate $%
t $ it is convenient to introduce the conformal time $\tau $ in accordance
with%
\begin{equation}
\tau =-\alpha e^{-t/\alpha },\;-\infty <\tau <0.  \label{eta}
\end{equation}%
In terms of this coordinate the line element takes the conformally flat form:%
\begin{equation}
ds^{2}=(\alpha /\tau )^{2}[d\tau ^{2}-\sum_{i=1}^{3}(dz^{i})^{2}].
\label{ds2Dd}
\end{equation}

One of the characteristic features of field theory on backgrounds with
non-trivial topology is the appearance of inequivalent types of fields with
the same spin \cite{Isha78}. In particular, for fermion fields the boundary
conditions along the compactified dimensions can be either periodic
(untwisted field) or antiperiodic (twisted field). In this section we
consider the field with periodicity conditions (no summation over $l$):%
\begin{equation}
\psi (t,\mathbf{z}_{p},\mathbf{z}_{q}+L_{l}\mathbf{e}_{l})=\psi (t,\mathbf{z}%
_{p},\mathbf{z}_{q}),  \label{bc}
\end{equation}%
where $\mathbf{z}_{p}=(z^{1},\ldots ,z^{p})$ and $\mathbf{z}%
_{q}=(z^{p+1},\ldots ,z^{3})$ are the position vectors along uncompactified
and compactified dimensions, $\mathbf{e}_{l}$\ is the unit vector in the
direction of the coordinate $z^{l}$ with the length $L_{l}$, $0\leqslant
z^{l}\leqslant L_{l}$. The case of a fermionic field with antiperiodicity
conditions will be discussed in section \ref{sec:AntiPer}. The
compactification of the spatial dimensions leads to the modification of the
spectrum for zero point fluctuations of the fermionic field and as a result
to change of the vacuum expectation values (VEVs) of physical observables.
This is the well-known topological Casimir effect.

Among the most important quantities describing both local and global
properties of the vacuum are the fermionic condensate and the VEV of the
energy-momentum tensor. In order to evaluate these VEVs we expand the field
operator in terms of a complete set of positive and negative frequency
eigenspinors $\{\psi _{\beta }^{(+)},\psi _{\beta }^{(-)}\}$:
\begin{equation}
\hat{\psi}=\sum_{\beta }[\hat{a}_{\beta }\psi _{\beta }^{(+)}+\hat{b}_{\beta
}^{+}\psi _{\beta }^{(-)}],  \label{operatorexp}
\end{equation}%
where $\hat{a}_{\beta }$ is the annihilation operator for particles, and $%
\hat{b}_{\beta }^{+}$ is the creation operator for antiparticles. The
collective index $\beta $ specifies the eigenfunctions. Next, we substitute
expansion (\ref{operatorexp}) and the similar expansion for the Dirac
adjoint operator $\hat{\bar{\psi}}=\hat{\psi}^{+}\gamma ^{0}$ into the
expressions for the bilinear product $\hat{\bar{\psi}}\hat{\psi}$ and the
energy-momentum tensor,
\begin{equation}
T_{\mu \nu }\{\hat{\bar{\psi}},\hat{\psi}\}=\frac{i}{2}[\hat{\bar{\psi}}%
\gamma _{(\mu }\nabla _{\nu )}\hat{\psi}-(\nabla _{(\mu }\hat{\bar{\psi}}%
)\gamma _{\nu )}\hat{\psi}]\ .  \label{EMTform}
\end{equation}%
By making use of the standard anticommutation relations for the annihilation
and creation operators, one finds the following mode-sum formulae
\begin{eqnarray}
\langle 0|\bar{\psi}\psi |0\rangle &=&\sum_{\beta }\bar{\psi}_{\beta
}^{(-)}(x)\psi _{\beta }^{(-)}(x),  \label{modesum} \\
\langle 0|T_{\mu \nu }|0\rangle &=&\sum_{\beta }T_{\mu \nu }\{\bar{\psi}%
_{\beta }^{(-)}(x),\psi _{\beta }^{(-)}(x)\}\ ,  \label{modesumEMT}
\end{eqnarray}%
where $|0\rangle $ is the amplitude for the vacuum state.

In order to find the form of the fermionic eigenfunctions for the geometry
under consideration, we choose the basis tetrad in the form
\begin{equation}
e_{\mu }^{(0)}=\delta _{\mu }^{0},\;e_{\mu }^{(a)}=e^{t/\alpha }\delta _{\mu
}^{a},\;a=1,2,3.  \label{e}
\end{equation}%
By using Eq. (\ref{e}), for the components of the spin connection and for
the combination appearing in the Dirac equation we find
\begin{eqnarray}
\Gamma _{l} &=&\frac{e^{t/\alpha }}{2\alpha }\gamma ^{(0)}\gamma
^{(l)},\;l=1,2,3,  \notag \\
\Gamma _{0} &=&0,\;\gamma ^{\mu }\Gamma _{\mu }=\frac{3}{2\alpha }\gamma
^{(0)}.  \label{Gmu}
\end{eqnarray}%
In accordance with the symmetry of the problem under consideration the
spatial part of the eigenfunctions can be taken in the standard plane wave
form $e^{\pm i\mathbf{kr}}$, where $\mathbf{k}$ is the wave vector and the
upper/lower sign corresponds to the positive/negative frequency solutions.
We will decompose the wave vector into the components along the
uncomcatified and compactified dimensions, $\mathbf{k=}(\mathbf{k}_{p},%
\mathbf{k}_{q})$. For a spinor field with periodicity conditions along the
compactified dimensions one has%
\begin{equation}
\mathbf{k}_{q}=(2\pi n_{p+1}/L_{p+1},\ldots ,2\pi n_{3}/L_{3}),  \label{kq}
\end{equation}%
with $n_{p+1},\ldots ,n_{3}=0,\pm 1,\pm 2,\ldots $.

By taking into account formulae (\ref{Gmu}), from equation (\ref{Direq}) for
the positive frequency solutions one finds%
\begin{equation}
\left[ \gamma ^{(0)}\partial _{t}+ie^{-t/\alpha }(\mathbf{k}\cdot %
\boldsymbol{\gamma })+3\gamma ^{(0)}/2\alpha +im\right] \psi =0,  \label{De1}
\end{equation}%
where $\boldsymbol{\gamma }=(\gamma ^{1},\gamma ^{2},\gamma ^{2})$. Further,
we write the four-component spinor fields in terms of two-component ones as%
\begin{equation}
\psi =\left(
\begin{array}{c}
\varphi _{+} \\
\varphi _{-}%
\end{array}%
\right) .  \label{psidecomp}
\end{equation}%
From equation (\ref{De1}) one finds the set of two first order differential
equations for the functions $\varphi _{\pm }$:%
\begin{equation}
\left( \eta \partial _{\eta }-3/2\mp i\alpha m\right) \varphi _{\pm }-i\eta (%
\mathbf{k}\cdot \boldsymbol{\sigma })\varphi _{\mp }=0,  \label{xieq2}
\end{equation}%
where $\boldsymbol{\sigma }=(\sigma _{1},\sigma _{2},\sigma _{3})$ and we
have introduced a new independent variable $\eta $ in accordance with%
\begin{equation}
\eta =\alpha e^{-t/\alpha },\;0\leqslant \eta <\infty .  \label{etavar}
\end{equation}%
In accordance with (\ref{eta}) one has the simple relation $\eta =-\tau $
between this variable and the conformal time (note that we are still working
in the coordinate system defined by Eq. (\ref{ds2deSit})). Eqs. (\ref{xieq2}%
) lead to the following second order differential equations for the separate
functions:%
\begin{equation}
\left( \eta ^{2}\partial _{\eta }^{2}-3\eta \partial _{\eta }+k^{2}\eta
^{2}+\alpha ^{2}m^{2}\pm i\alpha m+15/4\right) \varphi _{\pm }=0,
\label{xieq4}
\end{equation}%
where%
\begin{equation}
k=|\mathbf{k}|=\sqrt{\mathbf{k}_{p}^{2}+\mathbf{k}_{q}^{2}}.  \label{ka}
\end{equation}

The solutions of equations (\ref{xieq4}) for the functions $\varphi _{\pm }$
are linear combinations of the functions $\eta ^{2}H_{\pm 1/2-i\alpha
m}^{(1)}(k\eta )$ and $\eta ^{2}H_{\pm 1/2-i\alpha m}^{(2)}(k\eta )$, with $%
H_{\nu }^{(1,2)}(x)$ being the Hankel functions. Different choices of the
coefficients in these linear combinations correspond to different choices of
the vacuum state. We will consider de Sitter invariant Bunch-Davies vacuum
\cite{Bunc78} for which the coefficients of the part containing the
functions $H_{\pm 1/2-i\alpha m}^{(2)}(k\eta )$ are zero. Hence, for a
fermionic field in the Bunch-Davies vacuum the solutions to Eqs. (\ref{xieq4}%
) are the functions%
\begin{equation}
\varphi _{+}=\varphi ^{(c)}\eta ^{2}H_{1/2-i\alpha m}^{(1)}(k\eta
),\;\varphi _{-}=-i\varphi ^{(c)}(\mathbf{n}\cdot \boldsymbol{\sigma })\eta
^{2}H_{-1/2-i\alpha m}^{(1)}(k\eta ),  \label{phixiH2}
\end{equation}%
where $\varphi ^{(c)}$ is an arbitrary constant spinor, $\mathbf{n}=\mathbf{k%
}/k$, and we have used Eq. (\ref{xieq2}) with the upper sign to express the
function $\varphi _{-}$ through the function $\varphi _{+}$.

On the base of solutions (\ref{phixiH2}) we can construct the positive
frequency solutions to the Dirac equation in the form (for the analog flat
spacetime construction see, for instance, \cite{Bere82})%
\begin{equation}
\psi _{\beta }^{(+)}=A_{\beta }\eta ^{2}e^{i\mathbf{k}\cdot \mathbf{r}%
}\left(
\begin{array}{c}
H_{1/2-i\alpha m}^{(1)}(k\eta )w^{(\sigma )} \\
-i(\mathbf{n}\cdot \boldsymbol{\sigma })H_{-1/2-i\alpha m}^{(1)}(k\eta
)w^{(\sigma )}%
\end{array}%
\right) ,  \label{psibet+n}
\end{equation}%
where $\sigma =\pm 1/2$, $\beta =(\mathbf{k},\sigma )$, and
\begin{equation}
w^{(1/2)}=\left(
\begin{array}{c}
1 \\
0%
\end{array}%
\right) ,\;w^{(-1/2)}=\left(
\begin{array}{c}
0 \\
1%
\end{array}%
\right) .  \label{wsig}
\end{equation}%
In the similar way, for the negative frequency solutions we find
\begin{equation}
\psi _{\beta }^{(-)}=A_{\beta }\eta ^{2}e^{-i\mathbf{k}\cdot \mathbf{r}%
}\left(
\begin{array}{c}
i(\mathbf{n}\cdot \boldsymbol{\sigma })H_{-1/2+i\alpha m}^{(2)}(k\eta
)w^{(\sigma )\prime } \\
H_{1/2+i\alpha m}^{(2)}(k\eta )w^{(\sigma )\prime }%
\end{array}%
\right) ,  \label{psibet-}
\end{equation}%
with $w^{(\sigma )^{\prime }}=2i\sigma w^{(\sigma )}$.

The normalization coefficient $A_{\beta }$ is determined from the
orthonormalization condition%
\begin{equation}
\int d^{3}x\,\sqrt{\gamma }\psi _{\beta }^{(\lambda )+}\psi _{\beta ^{\prime
}}^{(\lambda ^{\prime })}=\delta _{\beta \beta ^{\prime }}\delta _{\lambda
\lambda ^{\prime }},\;\lambda ,\lambda ^{\prime }=\pm ,  \label{normaliz}
\end{equation}%
with $\gamma $ being the determinant of the spatial metric. On the right of
this condition $\delta _{\beta \beta ^{\prime }}$ is understood as the Dirac
delta function for continuous indices and the Kronecker delta for discrete
ones. By making use of the Wronskian relation for the Hankel functions, one
finds%
\begin{equation}
A_{\beta }^{2}=\frac{ke^{\pi \alpha m}}{2^{p+2}\pi ^{p-1}V_{q}\alpha ^{3}},
\label{Abet}
\end{equation}%
where $V_{q}=L_{p+1}\cdots L_{3}$ is the volume of the compactified
subspace. For a massless fermionic field we have the standard conformal
relation $\psi _{\beta }^{(\pm )}=(\eta /\alpha )^{3/2}\psi _{\beta }^{%
\mathrm{(M)}(\pm )}$ between eigenspinors (\ref{psibet+n}), (\ref{psibet-})
defining the Bunch-Davies vacuum in dS spacetime and the corresponding
eigenspinors $\psi _{\beta }^{\mathrm{(M)}(\pm )}$ for the Minkowski
spacetime with spatial topology $\mathrm{R}^{p}\times (\mathrm{S}^{1})^{q}$.
Note that the plane wave eigenspinors of the type (\ref{psibet+n}) and (\ref%
{psibet-}) in uncompactified dS spacetime have been considered recently in
Ref. \cite{Cota02}.

\section{Fermionic condensate}

\label{sec:FermCond}

Substituting the eigenfunctions (\ref{psibet-}) into the mode-sum formula (%
\ref{modesum}), for the fermionic condensate in dS spacetime with spatial
topology $\mathrm{R}^{p}\times (\mathrm{S}^{1})^{q}$ one finds

\begin{eqnarray}
\langle \bar{\psi}\psi \rangle _{p,q} &=&\frac{\eta ^{4}e^{\pi \alpha m}}{%
2^{p}\pi ^{p/2-1}\Gamma (p/2)V_{q}\alpha ^{3}}\int_{0}^{\infty
}dk_{p}\,k_{p}^{p-1}\sum_{n_{\mathbf{q}}=-\infty }^{+\infty }k  \notag \\
&&\times \left[ H_{-1/2-i\alpha m}^{(1)}(k\eta )H_{-1/2+i\alpha
m}^{(2)}(k\eta )-H_{1/2-i\alpha m}^{(1)}(k\eta )H_{1/2+i\alpha
m}^{(2)}(k\eta )\right] ,  \label{condpq}
\end{eqnarray}%
where%
\begin{equation}
\sum_{n_{\mathbf{q}}=-\infty }^{+\infty }=\sum_{n_{p+1}=-\infty }^{+\infty
}\cdots \sum_{n_{3}=-\infty }^{+\infty }.  \label{nqsum}
\end{equation}%
The VEV given by formula (\ref{condpq}) is divergent and needs some
regularization procedure. To make it finite we can introduce a cut-off
function $\varphi _{\lambda }(k)$ with the cut-off parameter $\lambda $,
which decreases sufficiently fast with increasing $k$ and satisfies the
condition $\varphi _{\lambda }(k)\rightarrow 1$, for $\lambda \rightarrow 0$%
. As a next step we apply to the series over $n_{p+1}$ the Abel-Plana
summation formula \cite{Most97,Saha07Gen}%
\begin{equation}
\sideset{}{'}{\sum}_{n=0}^{\infty }f(n)=\int_{0}^{\infty
}dx\,f(x)+i\int_{0}^{\infty }dx\,\frac{f(ix)-f(-ix)}{e^{2\pi x}-1},
\label{Abel}
\end{equation}%
where the prime means that the term $n=0$ should be halved. The term in the
VEV with the first integral in the right hand side of Eq. (\ref{Abel})
corresponds to the fermionic condensate in dS spacetime with topology $%
\mathrm{R}^{p+1}\times (\mathrm{S}^{1})^{q-1}$. As a result, the condensate
is presented in the decomposed form%
\begin{equation}
\langle \bar{\psi}\psi \rangle _{p,q}=\langle \bar{\psi}\psi \rangle
_{p+1,q-1}+\Delta _{p+1}\langle \bar{\psi}\psi \rangle _{p,q},
\label{fermconddec}
\end{equation}%
where the term%
\begin{eqnarray}
\Delta _{p+1}\langle \bar{\psi}\psi \rangle _{p,q} &=&\frac{2^{4-p}\eta
^{4}\alpha ^{-3}}{\pi ^{p/2+1}\Gamma (p/2)V_{q-1}}\int_{0}^{\infty
}dk_{p}\,k_{p}^{p-1}\sum_{n_{\mathbf{q-1}}=-\infty }^{+\infty
}\int_{0}^{\infty }dx\,x^{2}  \notag \\
&&\times \frac{{\mathrm{Im}}\left[ K_{1/2-i\alpha m}(\eta x)I_{1/2+i\alpha
m}(\eta x)\right] }{\sqrt{x^{2}+k_{p}^{2}+k_{\mathbf{n}_{q-1}}^{2}}%
(e^{L_{p+1}\sqrt{x^{2}+k_{p}^{2}+k_{\mathbf{n}_{q-1}}^{2}}}-1)},
\label{DeltCond}
\end{eqnarray}%
is induced by the compactness of the $(p+1)$th dimension. In Eq. (\ref%
{DeltCond}), $I_{\nu }(x)$ and $K_{\nu }(x)$ are the Bessel modified
functions, $V_{q-1}=L_{p+2}\cdots L_{3}$, and%
\begin{equation}
k_{\mathbf{n}_{q-1}}^{2}=\sum_{l=p+2}^{3}(2\pi n_{l}/L_{l})^{2}.
\label{kndmin1}
\end{equation}%
Note that the expression on the right of formula (\ref{DeltCond}) is finite
and we have safely removed the cut-off function. Of course, we could expect
the finiteness of the topological part as the toroidal compactification does
not change the local geometry and, hence, the structure of the divergences
is the same as in uncompactified dS spacetime.

An alternative form for the topological part of the fermionic condensate is
obtained expanding the function $1/(e^{y}-1)$ in the integrand and
explicitly integrating over $k_{p}$. This leads to the result%
\begin{eqnarray}
\Delta _{p+1}\langle \bar{\psi}\psi \rangle _{p,q} &=&\frac{32\eta
^{4}\alpha ^{-3}}{(2\pi )^{(p+3)/2}V_{q-1}}\sum_{n=1}^{\infty }\sum_{n_{%
\mathbf{q-1}}=-\infty }^{+\infty }\int_{0}^{\infty }dx\,x^{2}  \notag \\
&&\times \frac{{\mathrm{Im}}\left[ K_{1/2-i\alpha m}(\eta x)I_{1/2+i\alpha
m}(\eta x)\right] }{(nL_{p+1})^{p-1}}f_{(p-1)/2}(nL_{p+1}\sqrt{x^{2}+k_{%
\mathbf{n}_{q-1}}^{2}}),  \label{DeltCond2}
\end{eqnarray}%
where we have introduced the notation%
\begin{equation}
f_{\nu }(x)=x^{\nu }K_{\nu }(x).  \label{fnunot}
\end{equation}%
As it follows from formula (\ref{DeltCond2}), for a massless spinor field
the topological part of the fermionic condensate vanishes. From this formula
we also see that the topological part depends on the variable $\eta $ and
the length scales $L_{l}$ in the combinations $L_{l}/\eta $. Noting that $%
a(\eta )L_{l}$ is the comoving length with $a(\eta )=\alpha /\eta $ being
the scale factor, we conclude that the topological part of the fermionic
condensate is a function of comoving lengths of the compactified dimensions.
When the length of one of the compactified dimensions is large, $%
L_{l}\rightarrow \infty $, $l\geqslant p+2$, the main contribution into the
sum over $n_{l}$ in (\ref{DeltCond2}) comes from large values of $n_{l}$ and
we can replace the summation by the integration. The corresponding integral
can be evaluated by using the formula \cite{Prud86}
\begin{equation}
\int_{0}^{\infty }dz\,f_{(p-1)/2}(a\sqrt{z^{2}+b^{2}})=\sqrt{\frac{\pi }{2}}%
\frac{f_{p/2}(ab)}{a},  \label{IntegFormu}
\end{equation}%
and from Eq. (\ref{DeltCond2}) the formula is obtained for topology $\mathrm{%
R}^{p+1}\times (\mathrm{S}^{1})^{q-1}$.

On the base of the recurrence relation (\ref{fermconddec}) the fermionic
condensate can be decomposed as%
\begin{equation}
\langle \bar{\psi}\psi \rangle _{p,q}=\langle \bar{\psi}\psi \rangle _{%
\mathrm{dS}}+\langle \bar{\psi}\psi \rangle _{c},  \label{ConddS}
\end{equation}%
where the first term on the right is the condensate in uncompactified dS
spacetime and the part%
\begin{equation}
\langle \bar{\psi}\psi \rangle _{c}=\sum_{l=1}^{q}\Delta _{4-l}\langle \bar{%
\psi}\psi \rangle _{3-l,l},  \label{CondTop}
\end{equation}%
is induced by the non-trivial spatial topology.

Now we turn to the investigation of the topological part in the asymptotic
regions of the parameters. First let us consider the limit $L_{p+1}/\eta \ll
1$. This corresponds to the limit when the comoving length of the $(p+1)$th
direction is much smaller than the dS curvature radius: $a(\eta )L_{p+1}\ll
\alpha $. Introducing a new integration variable $y=L_{p+1}x$ and by taking
into account that for large values of $u$ one has%
\begin{equation}
{\mathrm{Im}}[I_{1/2+i\alpha m}(u)K_{1/2-i\alpha m}(u)]\sim -\frac{\alpha m}{%
2u^{2}},  \label{ImIK}
\end{equation}%
for the leading term in the asymptotic expansion of the fermionic condensate
we find%
\begin{equation}
\Delta _{p+1}\langle \bar{\psi}\psi \rangle _{p,q}\approx -\frac{8(\eta
/\alpha )^{2}m}{(2\pi )^{p/2+1}L_{p+1}^{p-1}V_{q}}\sum_{n=1}^{\infty
}\sum_{n_{\mathbf{q-1}}=-\infty }^{+\infty }\frac{f_{p/2}(nL_{p+1}k_{\mathbf{%
n}_{q-1}})}{n^{p}},\;L_{p+1}/\eta \ll 1.  \label{CondEarly}
\end{equation}%
Hence, in the limit under consideration the topological part is negative. As
due to the maximal symmetry of the dS spacetime the part $\langle \bar{\psi}%
\psi \rangle _{\mathrm{dS}}$ in the fermionic condensate is
time-independent, we conclude that in this limit the topological part
dominates. Taking into account the relation between $\eta $ and the
synchronous time coordinate, we see that formula (\ref{CondEarly}) describes
the asymptotic behavior in the early stages of the cosmological expansion
corresponding to $t\rightarrow -\infty $.

In the opposite limit, when $\eta /L_{p+1}\ll 1$, we again introduce the new
integration variable $y=L_{p+1}x$. By making use of the formulae for the
Bessel modified functions for small values of the arguments, to the leading
order we obtain%
\begin{equation}
\Delta _{p+1}\langle \bar{\psi}\psi \rangle _{p,q}\approx -\frac{%
2^{3-p/2}\alpha B_{0}e^{-4t/\alpha }}{\pi ^{(p+1)/2}L_{p+1}^{p+1}V_{q}\cosh
(\alpha m\pi )}\sin [2mt-2\alpha m\ln (\alpha /L_{p+1})-\phi _{0}],
\label{CondLate}
\end{equation}%
where $B_{0}$ and $\phi _{0}$ are defined by the relation%
\begin{equation}
B_{0}e^{i\phi _{0}}=\frac{2^{-i\alpha m}}{\Gamma (1/2+i\alpha m)}%
\sum_{n=1}^{\infty }\sum_{n_{\mathbf{q-1}}=-\infty }^{+\infty }\frac{%
f_{p/2+1+i\alpha m}(nL_{p+1}k_{\mathbf{n}_{q-1}})}{n^{p+2+2i\alpha m}}.
\label{B0}
\end{equation}%
In terms of the synchronous time coordinate this limit corresponds to the
late stages of the cosmological evolution, $t\rightarrow +\infty $. Hence,
in the limit when the comoving length of the compactified dimensions is much
larger than the curvature radius of dS spacetime, for a massive fermionic
field the topological part oscillates with exponentially decreasing
amplitude with respect to the synchronous time coordinate. Note that the
damping factor in the amplitude and the oscillation frequency are the same
for all terms in the sum on the left hand side of formula (\ref{CondTop})
and, hence, we have the similar oscillating behavior for the total
topological term: $\langle \bar{\psi}\psi \rangle _{c}\propto e^{-4t/\alpha
}\sin \left( 2mt+\phi _{c}\right) $.

\begin{figure}[tbph]
\begin{center}
\begin{tabular}{cc}
\epsfig{figure=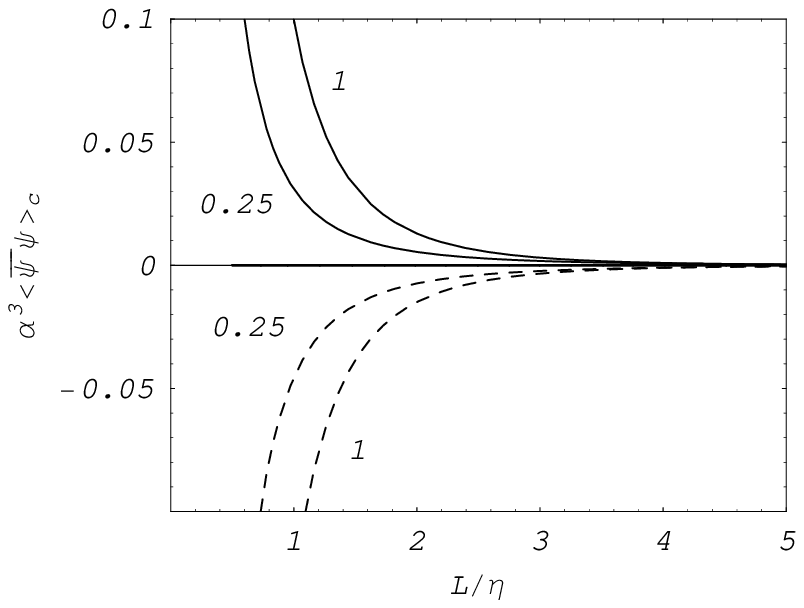,width=7.cm,height=6.cm} & \quad %
\epsfig{figure=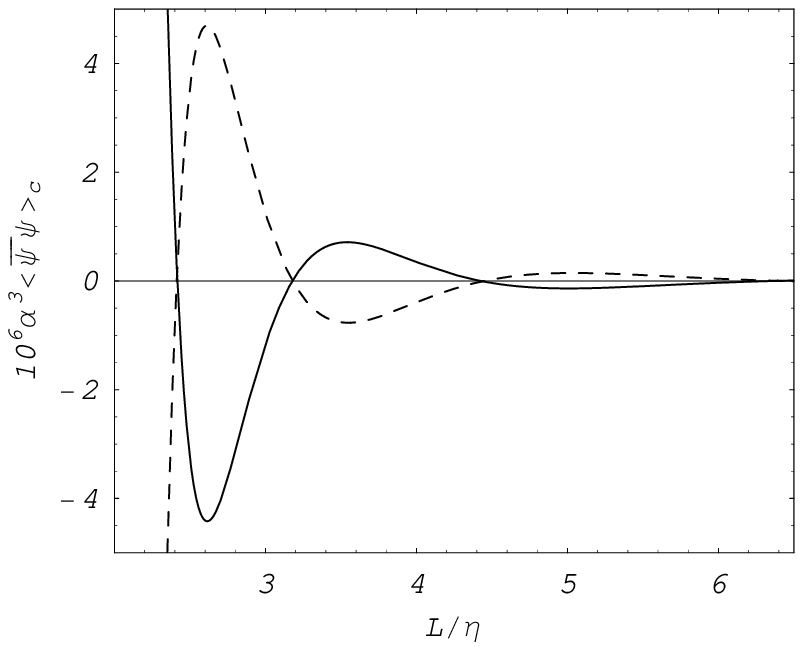,width=7.cm,height=6cm}%
\end{tabular}%
\end{center}
\caption{The topological part in the fermionic condensate for periodic
(dashed curves) and antiperiodic (full curves) spinor fields in dS spacetime
with spatial topology $\mathrm{R}^{2}\times \mathrm{S}^{1}$ as a function of
$L/\protect\eta =Le^{t/\protect\alpha }/\protect\alpha $. The numbers near
the curves on the left panel correspond to the values of the parameter $%
\protect\alpha m$. The graphs on the right panel are plotted for $\protect%
\alpha m=4$. }
\label{fig1}
\end{figure}

In the special case of spatial topology $\mathrm{R}^{2}\times \mathrm{S}^{1}$%
, from general formula (\ref{DeltCond2}) for the topological part in the
fermionic condensate one finds%
\begin{equation}
\langle \bar{\psi}\psi \rangle _{c}=-\frac{4\eta \alpha ^{-3}}{\pi ^{2}L}%
\int_{0}^{\infty }dx\,x^{2}{\mathrm{Im}}\left[ K_{1/2-i\alpha
m}(x)I_{1/2+i\alpha m}(x)\right] \ln (1-e^{-Lx/\eta }),  \label{R2S1}
\end{equation}%
where $L\equiv L_{3}$ is the length of the compactified dimension. On the
left panel of figure \ref{fig1} we have plotted this quantity for an
untwisted fermionic field (dashed curves) as a function of $L/\eta $ for the
values of the parameter $\alpha m=0.5,1$ (the numbers near the curves).
Writing $L/\eta =a(\eta )L/\alpha $, we see that this ratio is the comoving
length of the compactified dimension in units of the dS curvature radius.
As it follows from the asymptotic formula (\ref{CondLate}), for large values
of the ratio $L/\eta $ the topological part behaves like $\langle \bar{\psi}%
\psi \rangle _{c}\propto (\eta /L)^{4}\sin [2m\alpha \ln (\eta /L)+\phi _{0}]
$. As in the scale of the left panel the oscillations are not well seen, we
illustrate this oscillatory behavior on the right panel of figure \ref{fig1}%
, where the topological part in the fermionic condensate is plotted for an
untwisted fermionic field by the dashed curve versus $L/\eta $ for $\alpha
m=4$. The first zero with respect to $L/\eta $ and the distance between
neighbor zeros decrease with increasing values of the parameter $\alpha m$.

\section{Energy-momentum tensor}

\label{sec:EMT}

Now we turn to the investigation of the one-loop topological effects in the
VEV of the energy-momentum tensor for a fermionic field with periodic
boundary conditions. Substituting the eigenfunctions (\ref{psibet-}) into
the mode-sum formula (\ref{modesumEMT}), for the energy density and vacuum
stresses we find (no summation over $l$)%
\begin{eqnarray}
\langle 0|T_{0}^{0}|0\rangle  &=&\frac{2^{-p}\eta ^{5}\alpha ^{-4}}{\pi
^{p/2-1}\Gamma (p/2)V_{q}}\int_{0}^{\infty }dk_{p}\,k_{p}^{p-1}\sum_{n_{%
\mathbf{q}}=-\infty }^{+\infty }k^{2}  \notag \\
&&\times {\mathrm{Re}}\left[ H_{1/2+i\alpha m}^{(1)}(k\eta )H_{-1/2+i\alpha
m}^{(2)\prime }(k\eta )-H_{-1/2+i\alpha m}^{(1)}(k\eta )H_{1/2+i\alpha
m}^{(2)\prime }(k\eta )\right] ,  \label{T00} \\
\langle 0|T_{l}^{l}|0\rangle  &=&\frac{2^{1-p}\eta ^{5}\alpha ^{-4}}{\pi
^{p/2-1}\Gamma (p/2)V_{q}}\int_{0}^{\infty }dk_{p}\,k_{p}^{p-1}\sum_{n_{%
\mathbf{q}}=-\infty }^{+\infty }k_{l}^{2}{\mathrm{Re}}\left[ H_{1/2+i\alpha
m}^{(1)}(k\eta )H_{1/2+i\alpha m}^{(2)}(k\eta )\right] ,  \label{Tll}
\end{eqnarray}%
with $l=1,2,3$, and the off-diagonal components vanish. For a massless
field, by using the expressions for the functions $H_{1/2}^{(1,2)}(x)$, we
have the formula%
\begin{equation}
\langle 0|T_{k}^{l}|0\rangle =\frac{2(\eta /\alpha )^{4}}{(2\pi )^{p}V_{q}}%
\int d\mathbf{k}_{p}\sum_{n_{\mathbf{q}}=-\infty }^{+\infty }\frac{1}{k}%
\mathrm{diag}(-k^{2},k_{1}^{2},k_{2}^{2},k_{3}^{2}).  \label{Tlkm0}
\end{equation}%
It can be easily seen that the expression on the right of this formula
coincides with the leading term in the expansion of $\langle
0|T_{k}^{l}|0\rangle $ for a massive field in the limit $\eta \rightarrow
\infty $. Note that this limit corresponds to the adiabatic limit of the
slow expansion (for a detailed discussion see \cite{Birr82}).

As for the case of the fermionic condensate, in order to have finite
expressions we will assume that in expressions (\ref{T00}), (\ref{Tll}) a
cut-off function is introduced. By using the Abel-Plana formula, the VEV of
the energy-momentum tensor is presented in the decomposed form%
\begin{equation}
\langle T_{k}^{l}\rangle _{p,q}=\langle T_{k}^{l}\rangle _{p+1,q-1}+\Delta
_{p+1}\langle T_{k}^{l}\rangle _{p,q}.  \label{TllDecomp}
\end{equation}%
In this formula, $\langle T_{k}^{l}\rangle _{p+1,q-1}$ is the VEV of the
energy-momentum tensor for the topology $\mathrm{R}^{p+1}\times (\mathrm{S}%
^{1})^{q-1}$ and the part $\Delta _{p+1}\langle T_{k}^{l}\rangle _{p,q}$ is
due to the compactness of the $(p+1)$th dimension. The latter is given by
the formula (no summation over $l$)%
\begin{eqnarray}
\Delta _{p+1}\langle T_{l}^{l}\rangle _{p,q} &=&\frac{2^{2-p}\eta ^{5}\alpha
^{-4}}{\pi ^{p/2}\Gamma (p/2)V_{q-1}}\int_{0}^{\infty
}dk_{p}k_{p}^{p-1}\sum_{n_{\mathbf{q-1}}=-\infty }^{+\infty
}\int_{0}^{\infty }dz\,\,  \notag \\
&&\times \frac{zh^{(l)}(z)F^{(l)}(\eta z)}{\sqrt{z^{2}+k_{p}^{2}+k_{\mathbf{n%
}_{q-1}}^{2}}(e^{L_{p+1}\sqrt{z^{2}+k_{p}^{2}+k_{\mathbf{n}_{q-1}}^{2}}}-1)},
\label{TopTlln}
\end{eqnarray}%
with the notations%
\begin{eqnarray}
h^{(0)}(z) &=&-z^{2},\;h^{(l)}(z)=k_{p}^{2}/p,\;l=1,\ldots ,p,  \notag \\
h^{(p+1)}(z) &=&-(z^{2}+k_{p}^{2}+k_{\mathbf{n}_{q-1}}^{2}),\;%
\;h^{(l)}(z)=k_{l}^{2},\;l=p+2,3.  \label{hl}
\end{eqnarray}%
In Eq. (\ref{TopTll}), to simplify the formulae we have introduced the
notations%
\begin{eqnarray}
F^{(0)}(y) &=&\frac{2}{\pi }{\mathrm{Re}}\left[ I_{1/2+i\alpha
m}(y)K_{1/2-i\alpha m}^{\prime }(y)-K_{1/2-i\alpha m}(y)I_{1/2+i\alpha
m}^{\prime }(y)\right] ,  \label{F0} \\
F^{(l)}(y) &=&\frac{{\mathrm{Re}}[I_{-1/2-i\alpha m}^{2}(y)-I_{1/2+i\alpha
m}^{2}(y)]}{\cosh (\alpha m\pi )},\;l=1,2,3.  \label{Fl}
\end{eqnarray}%
The topological parts are finite and we have removed the cut-off functions
in the corresponding expressions. Note that the following relations take
place between the functions $F^{(l)}(y)$ and the corresponding function
appearing in the formula for the fermionic condensate:%
\begin{eqnarray}
F^{(0)}(y) &=&-F^{(1)}(y)-\frac{4\alpha m}{\pi y}{\mathrm{Im}}\left[
K_{1/2-i\alpha m}(y)I_{1/2+i\alpha m}(y)\right] ,  \notag \\
\partial _{y}[yF^{(0)}(y)] &=&\frac{4\alpha m}{\pi y}{\mathrm{Im}}\left[
K_{1/2-i\alpha m}(y)I_{1/2+i\alpha m}(y)\right] .  \label{RelFly}
\end{eqnarray}

Similar to the case of the fermionic condensate, the alternative form for
the topological parts is obtained expanding the integrand in Eq. (\ref%
{TopTlln}) and explicitly evaluating the integral over $k_{p}$:%
\begin{eqnarray}
\Delta _{p+1}\langle T_{l}^{l}\rangle _{p,q} &=&\frac{4\eta ^{5}\alpha ^{-4}%
}{(2\pi )^{(p+1)/2}V_{q-1}}\sum_{n=1}^{\infty }\sum_{n_{\mathbf{q-1}%
}=-\infty }^{+\infty }\int_{0}^{\infty }dx\,x  \notag \\
&&\times \frac{F^{(l)}(\eta x)}{(L_{p+1}n)^{p+1}}f_{p}^{(l)}(nL_{p+1}\sqrt{%
x^{2}+k_{\mathbf{n}_{q-1}}^{2}}),  \label{TopTll}
\end{eqnarray}%
where%
\begin{eqnarray}
f_{p}^{(0)}(y) &=&-(nL_{p+1}x)^{2}f_{(p-1)/2}(y),  \notag \\
f_{p}^{(l)}(y) &=&f_{(p+1)/2}(y),\;l=1,\ldots ,p,  \notag \\
f_{p}^{(p+1)}(y) &=&-\left[ pf_{(p+1)/2}(y)+y^{2}f_{(p-1)/2}(y)\right] ,\;\;
\label{fpl} \\
f_{p}^{(l)}(y) &=&k_{l}^{2}(nL_{p+1})^{2}f_{(p-1)/2}(y),\;l=p+2,\ldots ,3.
\notag
\end{eqnarray}%
Note that we have the relation%
\begin{equation}
\sum_{l=1}^{3}f_{p}^{(l)}(y)=f_{p}^{(0)}(y).  \label{rel1}
\end{equation}%
By using Eqs. (\ref{RelFly}), (\ref{rel1}), we can see that the topological
parts satisfy the standard trace relation
\begin{equation}
\Delta _{p+1}\langle T_{l}^{l}\rangle _{p,q}=m\Delta _{p+1}\langle \bar{\psi}%
\psi \rangle _{p,q}.  \label{TraceRel}
\end{equation}%
As an additional check of our calculations we can also see that the
topological parts obey the continuity equation $\left( \Delta _{p+1}\langle
T_{l}^{k}\rangle _{p,q}\right) _{;k}=0$, which for the geometry under
consideration takes the form%
\begin{equation}
\left( \eta \partial _{\eta }-4\right) \Delta _{p+1}\langle T_{0}^{0}\rangle
_{p,q}+\Delta _{p+1}\langle T_{l}^{l}\rangle _{p,q}=0.  \label{ContEq}
\end{equation}%
This relation is proved by using formulae (\ref{RelFly}), (\ref{rel1}) and (%
\ref{TraceRel}).

From formulae (\ref{RelFly}) the relation  $yF^{(1)}(y)=-(y^{2}F^{(0)}(y))^{%
\prime }$ is easily obtained. By making use of this relation and integrating
by parts the expression (\ref{TopTll}) for $l=1$, it can be seen that one
has the relation (no summation over $l$)%
\begin{equation}
\Delta _{p+1}\langle T_{l}^{l}\rangle _{p,q}=\Delta _{p+1}\langle
T_{0}^{0}\rangle _{p,q},\;l=1,\ldots p,  \label{stressEn}
\end{equation}%
between the energy density and the vacuum stresses along the uncompactified
dimensions. In deriving (\ref{stressEn}) we have also used the formula $%
f_{\nu }^{\prime }(z)=-zf_{\nu -1}$ which simply follows from well-known
properties of the function $K_{\nu }(z)$. Hence, in the uncompactified
subspace the equation of state for the topological part of the
energy-momentum tensor is of the cosmological constant type. Note that the
topological parts are time-dependent and they break the dS symmetry.

After the repetitive application of the recurrence formula (\ref{TllDecomp}%
), the VEV of the energy-momentum tensor for the topology $\mathrm{R}%
^{p}\times (\mathrm{S}^{1})^{q}$ is presented in the form%
\begin{equation}
\langle T_{l}^{k}\rangle _{p,q}=\langle T_{l}^{k}\rangle _{\mathrm{dS}%
}+\langle T_{l}^{k}\rangle _{c},\;\langle T_{l}^{k}\rangle
_{c}=\sum_{l=1}^{q}\Delta _{4-l}\langle T_{l}^{k}\rangle _{3-l,l},
\label{TlkdS}
\end{equation}%
where $\langle T_{l}^{k}\rangle _{\mathrm{dS}}$ is the VEV in the
uncompactified dS spacetime and $\langle T_{l}^{k}\rangle _{c}$ is the
topological part. As the Bunch-Davies vacuum is dS invariant, the
energy-momentum tensor $\langle T_{l}^{k}\rangle _{\mathrm{dS}}$ corresponds
to a gravitational source of the cosmological constant type. In particular,
combining with the initial cosmological constant $\Lambda $, one-loop
effects in uncompactified dS spacetime lead to the effective cosmological
constant $\Lambda _{\mathrm{eff}}=\Lambda +8\pi G\langle T_{0}^{0}\rangle _{%
\mathrm{dS}}$, where $G$ is the Newton's gravitational constant.

For a massless fermionic field, by using the expressions for the functions $%
I_{\pm 1/2}(y)$, $K_{1/2}(y)$, we see that%
\begin{equation}
F^{(0)}(y)=-1/y,\;F^{(1)}(y)=2/\pi y,\;m=0.  \label{F01m0}
\end{equation}%
Now the integration over $x$ in (\ref{TopTll}) is done explicitly and one
finds (no summation over $l$)%
\begin{equation}
\Delta _{p+1}\langle T_{l}^{l}\rangle _{p,q}=\frac{8(\eta /\alpha )^{4}}{%
(2\pi )^{p/2+1}V_{q}L_{p+1}^{p+1}}\sum_{n=1}^{\infty }\sum_{\mathbf{n}%
_{q-1}=-\infty }^{+\infty }\frac{g_{p}^{(l)}(nL_{p+1}k_{\mathbf{n}_{q-1}})}{%
n^{p+2}},  \label{DelTConf}
\end{equation}%
with the notations%
\begin{eqnarray}
g_{p}^{(0)}(y) &=&g_{p}^{(l)}(y)=f_{p/2+1}(y),\;l=1,\ldots ,p,  \notag \\
g_{p}^{(p+1)}(y) &=&-(p+1)f_{p/2+1}(y)-y^{2}f_{p/2}(y),  \label{gi} \\
g_{p}^{(l)}(y) &=&(nL_{p+1}k_{l})^{2}f_{p/2}(y),\;l=p+2,\ldots ,3.  \notag
\end{eqnarray}%
Note that in this case the problem under consideration is conformally
related to the corresponding problem in the Minkowski spacetime with spatial
topology $\mathrm{R}^{p}\times (\mathrm{S}^{1})^{q}$ and formula (\ref%
{DelTConf}) is obtained from the relation $\Delta _{p+1}\langle
T_{i}^{i}\rangle _{p,q}=a^{-4}(\eta )\Delta _{p+1}\langle T_{i}^{i}\rangle
_{p,q}^{\mathrm{(M)}}$, with $a(\eta )=\alpha /\eta $ being the scale
factor. Of course, this relation is valid for general case of the scale
factor $a(\eta )$. Comparing expression (\ref{DelTConf}) with the
corresponding formula from \cite{Bell08} for a conformally coupled massless
scalar field, we see that the following relation takes place: $\Delta
_{p+1}\langle T_{k}^{l}\rangle _{p,q}=-4\Delta _{p+1}\langle
T_{k}^{l}\rangle _{p,q}^{\mathrm{(scalar)}}$. The latter is a simple
generalization of the corresponding topologically non-trivial flat spacetime
result (see, for instance, \cite{Birr79}). In curved backgrounds this
relation between the topological parts of the fermionic and scalar
energy-momentum tensors is not valid for a minimally coupled scalar field.

Let us consider the behavior of the vacuum energy-momentum tensor in the
asymptotic regions of the parameters. In the limit $L_{p+1}/\eta \ll 1$,
corresponding to small values of the comoving length $a(\eta )L_{p+1}$ with
respect to the dS curvature radius, we introduce a new integration variable $%
y=L_{p+1}x$ in Eq. (\ref{TopTll}) and expand the functions $F^{(l)}(\eta
y/L_{p+1})$. By taking into account that for large values $z$ to the leading
order%
\begin{equation}
F^{(0)}(z)\approx -1/z,\;F^{(1)}(z)\approx 2/\pi z,\;z\gg 1,
\label{F01LargeArg}
\end{equation}%
we see that in this order the topological part in the VEV of the
energy-momentum tensor coincides with that for a massless field given by
formula (\ref{DelTConf}). In particular, the topological part of the vacuum
energy density is negative. Note that the limit under consideration
corresponds to the early stages of the cosmological evolution, $t\rightarrow
-\infty $. In this limit the topological parts in the VEVs of the
energy-momentum tensors for periodic scalar and fermionic fields have
opposite signs.

For small values of the ratio $\eta /L_{p+1}$, when the comoving length $%
a(\eta )L_{p+1}$ is larger that the dS curvature radius, after introducing
the new integration variable $y=L_{p+1}x$, we see that the argument of the
functions $F^{(l)}(\eta y/L_{p+1})$ is small. By making use of the formulae
for the Bessel modified functions for small arguments, in the leading order
we obtain the formula (no summation over $l$)%
\begin{equation}
\Delta _{p+1}\langle T_{l}^{l}\rangle _{p,q}\approx \frac{%
2^{2-p/2}B_{l}e^{-4t/\alpha }}{\pi ^{(p+1)/2}V_{q}L_{p+1}^{p+1}\cosh (\alpha
m\pi )}\cos [2mt-2\alpha m\ln (\alpha /L_{p+1})-\phi _{l}].  \label{TllSmall}
\end{equation}%
Here $B_{0}$, $\phi _{0}$ are defined by relation (\ref{B0}), $B_{l}=B_{0}$,
$\phi _{l}=\phi _{0}$ for $l=1,\ldots ,p$, and $B_{l}$, $\phi _{l}$, $%
l=p+1,\ldots ,3$, are defined by%
\begin{eqnarray}
B_{p+1}e^{i\phi _{p+1}} &=&\frac{-2^{-i\alpha m}}{\Gamma (1/2+i\alpha m)}%
\sum_{n=1}^{\infty }\sum_{n_{\mathbf{q-1}}=-\infty }^{+\infty }\frac{1}{%
n^{p+2+2i\alpha m}}  \notag \\
&&\times \left[ (p+1+2i\alpha m)f_{p/2+1+i\alpha m}(x)+x^{2}f_{p/2+i\alpha
m}(x)\right] _{x=nL_{p+1}k_{\mathbf{n}_{q-1}}},  \label{Bp+1} \\
B_{l}e^{i\phi _{l}} &=&\frac{2^{-i\alpha m}}{\Gamma (1/2+i\alpha m)}%
\sum_{n=1}^{\infty }\sum_{n_{\mathbf{q-1}}=-\infty }^{+\infty
}(L_{p+1}k_{l})^{2}\frac{f_{p/2+i\alpha m}(nL_{p+1}k_{\mathbf{n}_{q-1}})}{%
n^{p+2i\alpha m}},  \label{Blphi}
\end{eqnarray}%
and in the last formula $l=p+2,\ldots ,3$. Note that if the lengths of the
compactified dimensions are the same the quantities $B_{l}$, $\phi _{l}$ do
not depend on the compactification length. Formula (\ref{TllSmall})
describes the asymptotic behavior of the topological part in the late stages
of cosmological evolution which in terms of the synchronous time coordinate
correspond to the limit $t\rightarrow +\infty $. As we see, in this limit
the behavior of the topological part in the VEV of the energy-momentum
tensor for a massive spinor field is damping oscillatory. This type of
oscillations are absent in the case of a massless field when the topological
parts behave like $e^{-4t/\alpha }$. For a massive field the damping factor
in the amplitude and the oscillation frequency are the same for all terms in
the sum of Eq. (\ref{TlkdS}) and the total topological term behaves like $%
\langle T_{l}^{k}\rangle _{c}\propto e^{-4t/\alpha }\cos \left( 2mt+\phi
_{c}^{\prime }\right) $. As the vacuum energy-momentum tensor for
uncompactified dS spacetime is time-independent, we have similar damping
oscillations in the total energy-momentum tensor $\langle T_{l}^{k}\rangle _{%
\mathrm{dS}}+\langle T_{l}^{k}\rangle _{c}$ as well.

The general formulae for the topological part in the VEV\ of the
energy-momentum tensor are further simplified in the special case of
topology $\mathrm{R}^{2}\times \mathrm{S}^{1}$. For this case the functions $%
f_{p}^{(l)}(x)$ are expressed in terms of exponentials and after the
summation over $n$ we find (no summation over $l$)%
\begin{equation}
\langle T_{l}^{l}\rangle _{c}=\frac{(\eta /L)^{3}}{\pi \alpha ^{4}}%
\int_{0}^{\infty }dx\,xF^{(l)}(x)G_{l}(Lx/\eta ),  \label{TllR2S1}
\end{equation}%
with $L=L_{3}$ being the length of the compactified dimension. In Eq. (\ref%
{TllR2S1}) the following notations are introduced%
\begin{eqnarray}
&&G_{0}(y)=y^{2}\ln (1-e^{-y}),  \notag \\
&&G_{1}(y)=G_{2}(y)=y\mathrm{Li}_{2}(e^{-y})+\mathrm{Li}_{3}(e^{-y}),
\label{Gly} \\
&& G_{3}(y)=G_{0}(y)-2G_{1}(y),  \notag
\end{eqnarray}%
and $\mathrm{Li}_{n}(z)=\sum_{k=1}^{\infty }z^{k}/k^{n}$ is the
polylogarithm function. For a massless fermionic field we have%
\begin{equation}
\langle T_{k}^{l}\rangle _{c}=\frac{2\pi ^{2}}{45}\left( \frac{\eta }{\alpha
L}\right) ^{4}\mathrm{diag}(1,1,1,-3).  \label{TlkR2S1m0}
\end{equation}

\begin{figure}[tbph]
\begin{center}
\begin{tabular}{cc}
\epsfig{figure=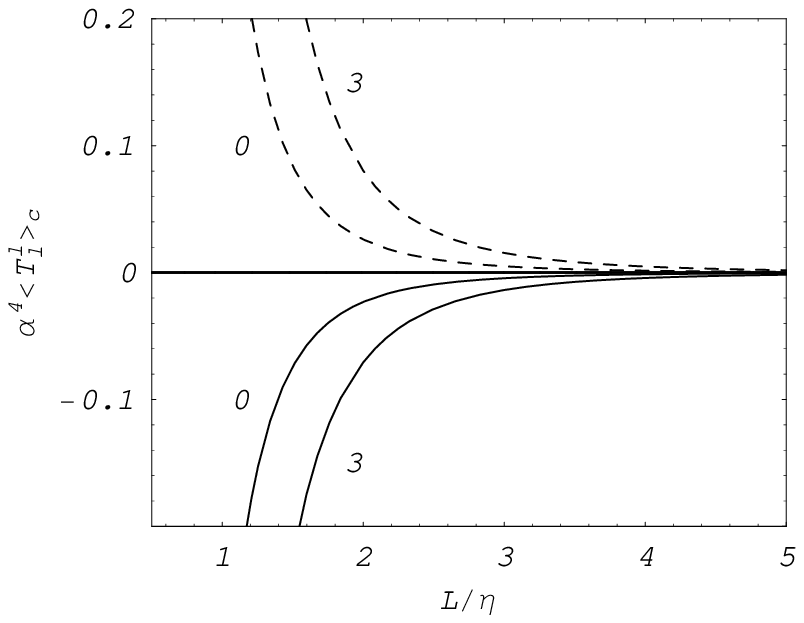,width=7.cm,height=6cm} & \quad %
\epsfig{figure=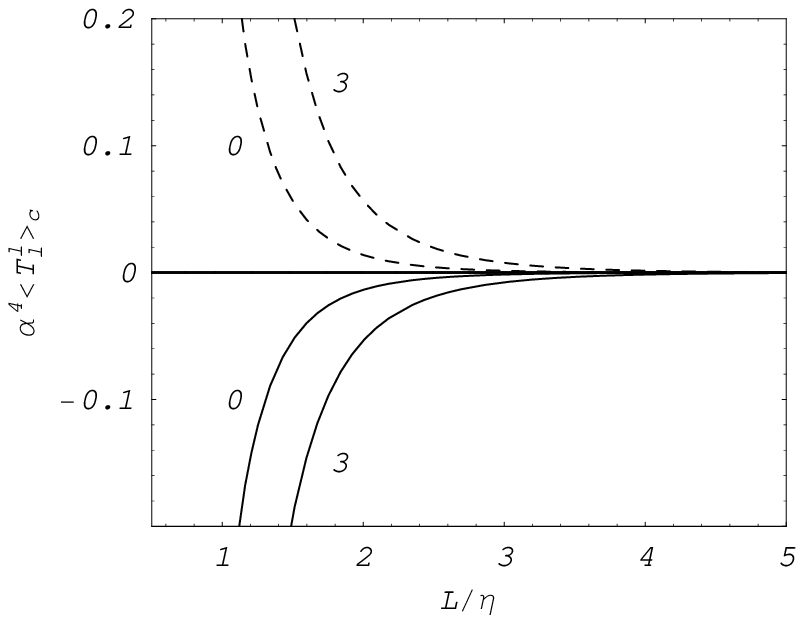,width=7.cm,height=6cm}%
\end{tabular}%
\end{center}
\caption{The topological parts in the VEV of the energy density
($l=0$) and the stress along the compactified dimension ($l=3$)
for periodic (dashed curves) and antiperiodic (full curves) spinor
fields in dS spacetime with spatial topology $\mathrm{R}^{2}\times
\mathrm{S}^{1}$ as functions of $L/\protect\eta
=Le^{t/\protect\alpha }/\protect\alpha $. The numbers near the
curves correspond to the values of the index $l$. The vacuum
stresses along the uncompactified dimensions are equal to the
energy density. For the left panel $\protect\alpha m=0.25$ and for
the right one $\protect\alpha m=1$. } \label{fig2}
\end{figure}

In figure \ref{fig2} by the dashed curves we have presented the dependence
of the topological part in the VEV\ of the energy-momentum tensor (\ref%
{TllR2S1}) on the ratio $L/\eta $ for an untwisted fermionic field in the
case of topology $\mathrm{R}^{2}\times \mathrm{S}^{1}$. The numbers near the
curves correspond to the values of the index $l$. Note that, as we have seen
above, the stresses along uncompactified dimensions ($l=1,2$) coincide with
the energy density.

\begin{figure}[tbph]
\begin{center}
\epsfig{figure=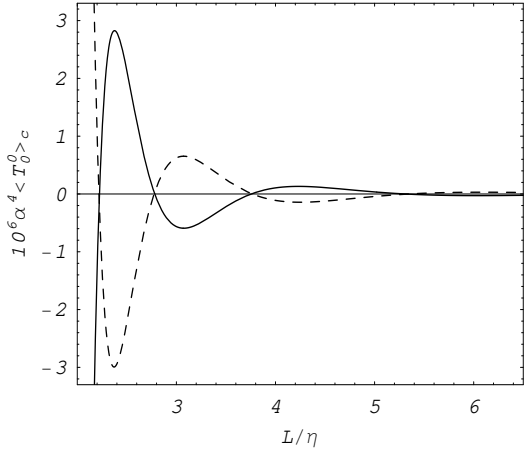,width=7.cm,height=6cm}
\end{center}
\caption{The topological parts in the VEV of the energy density for periodic
(dashed curve) and antiperiodic (full curves) spinor fields in dS spacetime
with spatial topology $\mathrm{R}^{2}\times \mathrm{S}^{1}$ as functions of $%
L/\protect\eta $. The graphs are plotted for the value of the parameter $%
\protect\alpha m=4$. }
\label{fig3}
\end{figure}

As it has been shown before, for a massive field the topological part
oscillates for large values of $L/\eta $. For an untwisted fermionic field
this is shown separately in figure \ref{fig3}, where the energy density ($l=0
$) is plotted by the dashed curve for the value of the parameter $\alpha m=4$%
. As in the case of the fermionic condensate, the first zero and the
distance between the neighbor zeros decrease with increasing $\alpha m$.

\section{Spinor field with antiperiodicity conditions}

\label{sec:AntiPer}

In this section we consider the spinor field satisfying the antiperiodicity
condition along the compactified dimensions%
\begin{equation}
\psi (t,\mathbf{z}_{p},\mathbf{z}_{q}+L_{l}\mathbf{e}_{l})=-\psi (t,\mathbf{z%
}_{p},\mathbf{z}_{q}).  \label{AntCond}
\end{equation}%
The corresponding eigenfunctions are given by formulae (\ref{psibet+n}), (%
\ref{psibet-}), where now the wave vector along the compactified diemnsions
is given by the formula
\begin{equation}
\mathbf{k}_{q}=(\pi (2n_{p+1}+1)/L_{p+1},\ldots ,\pi (2n_{3}+1)/L_{3}),
\label{kqCompTw}
\end{equation}%
and $n_{p+1},\ldots ,n_{3}=0,\pm 1,\pm 2,\ldots $. The fermionic condensate
and the VEV of the energy-momentum tensor are still given by formulae (\ref%
{T00}), (\ref{Tll}) with%
\begin{equation}
k^{2}=k_{p}^{2}+\sum_{l=p+1}^{3}\left[ \pi (2n_{l}+1)/L_{l}\right] ^{2}.
\label{kTw}
\end{equation}%
Further, we apply to the series over $n_{p+1}$ the Abel-Plana formula in the
form \cite{Most97,Saha07Gen}%
\begin{equation}
\sum_{n=0}^{\infty }f(n+1/2)=\int_{0}^{\infty }dx\,f(x)-i\int_{0}^{\infty
}dx\,\frac{f(ix)-f(-ix)}{e^{2\pi x}+1}.  \label{abel2}
\end{equation}%
As a result, the topological part in the fermionic condensate is presented
in the form (\ref{fermconddec}), where now the part induced by the
compactness of the $(p+1)$th direction is given by the formula%
\begin{eqnarray}
\Delta _{p+1}\langle \bar{\psi}\psi \rangle _{p,q} &=&-\frac{4\eta
^{4}\alpha ^{-3}}{2^{p-1}\pi ^{p+1}V_{q-1}}\int_{0}^{\infty
}dk_{p}\,k_{p}^{p-1}\sum_{n_{\mathbf{q-1}}=-\infty }^{+\infty
}\int_{0}^{\infty }dz\,  \notag \\
&&\times z^{2}\frac{{\mathrm{Im}}\left[ K_{1/2-i\alpha m}(\eta
z)I_{1/2+i\alpha m}(\eta z)\right] }{\sqrt{z^{2}+k_{p}^{2}+k_{\mathbf{n}%
_{q-1}}^{2}}(e^{L_{p+1}\sqrt{z^{2}+k_{p}^{2}+k_{\mathbf{n}_{q-1}}^{2}}}+1)}.
\label{FCAnty}
\end{eqnarray}%
with the notation
\begin{equation}
k_{\mathbf{n}_{q-1}}^{2}=\sum_{l=p+2}^{3}\left[ \pi (2n_{l}+1)/L_{l}\right]
^{2}.  \label{knq-1Tw}
\end{equation}%
Expanding the integrand and performing the integration over $k_{p}$, this
part can also be presented in the form%
\begin{eqnarray}
\Delta _{p+1}\langle \bar{\psi}\psi \rangle _{p,q} &=&\frac{32\eta
^{4}\alpha ^{-3}}{(2\pi )^{(p+3)/2}V_{q}L_{p+1}^{p-2}}\sum_{n=1}^{\infty
}(-1)^{n}\sum_{n_{\mathbf{q-1}}=-\infty }^{+\infty }\int_{0}^{\infty
}dx\,x^{2}  \notag \\
&&\times \frac{{\mathrm{Im}}\left[ K_{1/2-i\alpha m}(\eta x)I_{1/2+i\alpha
m}(\eta x)\right] }{n^{p-1}}f_{(p-1)/2}(nL_{p+1}\sqrt{x^{2}+k_{\mathbf{n}%
_{q-1}}^{2}}),  \label{FCAnty2}
\end{eqnarray}%
where the function $f_{\nu }(y)$ is defined by Eq. (\ref{fnunot}). For a
massless field the topological part vanishes. Having the parts induced by
separate compact dimensions we can decompose the fermionic condensate as
given in Eqs. (\ref{ConddS}), (\ref{CondTop}).

General formula (\ref{FCAnty2}) is simplified in the special case of
topology $\mathrm{R}^{2}\times \mathrm{S}^{1}$:%
\begin{equation}
\langle \bar{\psi}\psi \rangle _{c}=-\frac{4\eta \alpha ^{-3}}{\pi ^{2}L}%
\int_{0}^{\infty }dx\,x^{2}{\mathrm{Im}}\left[ K_{1/2-i\alpha
m}(x)I_{1/2+i\alpha m}(x)\right] \ln (1+e^{-Lx/\eta }),  \label{CondTwR2S1}
\end{equation}%
where $L$ is the length of the compact direction. On the left panel of
figure \ref{fig1} by full curves we have plotted the dependence of this
quantity on the ratio $L/\eta $. The numbers near the curves correspond to
the values of the parameter $\alpha m$. The full curve on the right panel of
figure \ref{fig1} shows the oscillatory behavior of the fermionic condensate
for a twisted field. The curve is plotted for the value $\alpha m=4$.

In the similar way, for the topological parts in the VEV of the
energy-momentum tensor we find the formula (no summation over $l$)%
\begin{eqnarray}
\Delta _{p+1}\langle T_{l}^{l}\rangle _{p,q} &=&\frac{4\eta ^{5}\alpha ^{-4}%
}{(2\pi )^{(p+1)/2}V_{q}L_{p+1}^{p}}\sum_{n=1}^{\infty }(-1)^{n}\sum_{n_{%
\mathbf{q-1}}=-\infty }^{+\infty }\int_{0}^{\infty }dx\,  \notag \\
&&\times x\frac{F^{(l)}(\eta x)}{n^{p+1}}f_{p}^{(l)}(nL_{p+1}\sqrt{x^{2}+k_{%
\mathbf{n}_{q-1}}^{2}}).  \label{TllTw}
\end{eqnarray}%
Hence, in the case of a fermionic field with antiperiodicity conditions the
formulae for the topological parts in the VEVs are obtained from those for
the field with periodicity conditions inserting the factor $(-1)^{n}$ in the
summation over $n$ and replacing the definition for $k_{\mathbf{n}%
_{q-1}}^{2} $ by (\ref{knq-1Tw}). Note that, as in the case of periodic
boundary conditions, we have relation (\ref{stressEn}) between the energy
density and the vacuum stresses along the uncompactified dimensions.

For the case of topology $\mathrm{R}^{2}\times \mathrm{S}^{1}$, in the way
similar to that for periodic field, the formulae (\ref{TllTw}) are
simplified giving (no summation over $l$)%
\begin{equation}
\langle T_{l}^{l}\rangle _{c}=\frac{(\eta /L)^{3}}{\pi \alpha ^{4}}%
\int_{0}^{\infty }dx\,xF^{(l)}(x)G_{l}^{\mathrm{(tw)}}(Lx/\eta ),
\label{TllTwR2S1}
\end{equation}%
with the notations%
\begin{eqnarray}
&&G_{0}^{\mathrm{(tw)}}(y)=y^{2}\ln (1+e^{-y}),  \notag \\
&&G_{l}^{\mathrm{(tw)}}(y)=y\mathrm{Li}_{2}(-e^{-y})+\mathrm{Li}%
_{3}(-e^{-y}),\;l=1,2,  \label{G0tw} \\
&&G_{3}^{\mathrm{(tw)}}(y)=G_{0}^{\mathrm{(tw)}}(y)-2G_{l}^{\mathrm{(tw)}%
}(y).  \notag
\end{eqnarray}%
For a massless field we have the result%
\begin{equation*}
\langle T_{k}^{l}\rangle _{c}=-\frac{7\pi ^{2}}{180}\left( \frac{\eta }{%
\alpha L}\right) ^{4}\mathrm{diag}(1,1,1,-3).
\end{equation*}%
In figure \ref{fig2} (full curves) the topological part in the VEV\ of the
energy-momentum tensor (\ref{TllTwR2S1}) for a twisted fermionic field is
plotted versus $L/\eta $. The numbers near the curves correspond to the
values of the index $l$. As it has been shown before, for a massive field
the topological parts oscillate for large values of $L/\eta $. The
oscillations of the topological part in the energy density as a function of $%
L/\eta $ are illustrated in figure \ref{fig3} by the full curve plotted for
the value of parameter $\alpha m=4$.

\section{Conclusion}

\label{sec:Conc}

Motivated by the importance of compactified dimensions in physical models
and continuing our work \cite{Saha07,Bell08}, in the present paper we have
investigated the fermionic condensate and the VEV of the energy-momentum
tensor for a massive fermionic field in dS spacetime with toroidally
compactified spatial dimensions. In order to evaluate the corresponding
mode-sums we need the eigenspinors satisfying appropriate boundary
conditions along the compactified dimensions. For a fermionic field with
periodicity conditions these eigenspinors are constructed in section \ref%
{sec:EigFunc} and are given by formulae (\ref{psibet+n}), (\ref{psibet-}).
Choosing these functions we have assumed that the field is in the
Bunch-Davies vacuum state. The application of the Abel-Plana formula allows
us to present the VEVs for the spatial topology $\mathrm{R}^{p}\times (%
\mathrm{S}^{1})^{q}$ as the sum of the corresponding quantity in the
topology $\mathrm{R}^{p+1}\times (\mathrm{S}^{1})^{q-1}$ and of the part
which is induced by the compactness of $(p+1)$th dimension. The latter is
finite and in this way the renormalization is reduced to that for the
uncompactified dS spacetime. In the case of an untwisted field the
topological parts in the fermionic condensate and in the VEV of the
energy-momentum tensor are given by formulae (\ref{DeltCond2}), (\ref{TopTll}%
). The parts induced by the non-trivial topology are time-dependent and
breake the dS symmetry. The corresponding vacuum stresses along the
uncompactified dimensions coincide with the energy density and, hence, in
the uncompactified subspace the equation of state for the topological part
of the energy-momentum tensor is of the cosmological constant type. As an
additional check of the formulae, we have shown that the topological parts
satisfy trace relation (\ref{TraceRel}) and the covariant continuity
equation (\ref{ContEq}). For a massless fermionic field the problem under
consideration is conformally related to the corresponding problem in the
Minkowski spacetime with spatial topology $\mathrm{R}^{p}\times (\mathrm{S}%
^{1})^{q}$ and the fermionic condensate vanishes. In this case we have the
relation $\langle T_{k}^{l}\rangle _{c}=a^{-4}(\eta )\langle
T_{k}^{l}\rangle _{c}^{\mathrm{(M)}}$ between the topological contributions
in the VEV of the energy-momentum tensor (see formula (\ref{DelTConf})). For
a massless field the topological part in the VEV of the fermionic
energy-momentum tensor differs from the corresponding quantity for a
conformally coupled massless scalar field by an additional factor -4.

For the case of a massive field the general formulae are simplified in the
asymptotic regions of the parameters. In the limit when the comoving length
of a compactified dimension is much smaller than the dS curvature radius, $%
L_{p+1}/\eta \ll 1$, for an untwisted field to the leading order
the topological part in the fermionic condensate is given by
formula (\ref{CondEarly}) and is negative. In the same limit, the
topological part in the VEV of the energy-momentum tensor
coincides with the corresponding quantity for a massless field and
is conformally related to the corresponding VEV in toroidally
compactified Minkowski spacetime. In particular, the vacuum energy
density is negative. This limit corresponds to the early stages of
the cosmological evolution and the topological parts dominate over
the uncompactified dS parts. In the opposite limit, when the
comoving lengths of the compactified dimensions are large with
respect to the dS curvature radius, $L_{p+1}/\eta \gg 1$, in the
case of a massive field the asymptotic behavior of the topological
parts is damping oscillatory for both fermionic condensate and the
energy-momentum tensor and to the leading order is given by
formulae (\ref{CondLate}), (\ref{TllSmall}). These formulae
describe the behavior of the topological parts in the late stages
of the cosmological expansion. As the uncompactified dS parts are
time-independent, we have similar oscillations in the total VEVs
as well. Note that this type of oscillatory behavior is absent for
a massless fermionic field.

In section \ref{sec:AntiPer} we have considered a fermionic field with
antiperiodicity conditions along the compactified dimensions. In this case
the topological parts in the fermionic condensate and the VEV of the
energy-momentum tensor are given by formulae (\ref{FCAnty2}), (\ref{TllTw}).
The asymptotic expressions for these VEVs in the early and late stages of
the cosmological expansion are obtained from the corresponding formulae for
an untwisted field inserting the factor $(-1)^{n}$ in the summation over $n$
and defining $k_{\mathbf{n}_{q-1}}^{2}$ in accordance with Eq. (\ref{knq-1Tw}%
). The features in the behavior of the topological VEVs we have illustrated
in figures \ref{fig1}-\ref{fig3} for both untwisted and twisted fermionic
fields in the case of spatial topology $\mathrm{R}^{2}\times \mathrm{S}^{1}$%
. For this topology the general expressions are further simplified to
formulae (\ref{R2S1}), (\ref{TllR2S1}), (\ref{CondTwR2S1}), (\ref{TllTwR2S1}%
).

\section{Acknowledgments}

I am grateful to Emilio Elizalde and Sergei Odintsov for their kind
hospitality during my stay at Instituci\`{o} Catalana de Recerca i Estudis
Avan\c{c}ats (ICREA) and Institut de Ci\`{e}ncies de l'Espai (IEEC-CSIC)
(Barcelona, Spain). The work was supported in part by the Armenian Ministry
of Education and Science Grant No. 119.


\begin{thebibliography}{99}
\bibitem{Lach95} M. Lachi\`{e}ze-Rey and J.-P. Luminet, Phys. Rep. \textbf{%
254}, 135 (1995).

\bibitem{Levi02} J. Levin, Phys. Rep. \textbf{365}, 251 (2002); N.J.
Cornish, D.N. Spergel, G.D. Starkman, and E. Komatsu, Phys. Rev. Lett.
\textbf{92}, 201302 (2004).

\bibitem{Lind04} A. Linde, JCAP \textbf{0410}, 004 (2004).

\bibitem{Star98} G.D. Starkman, Class. Quantum Grav. \textbf{15}, 2529
(1998); N.J. Cornish, D.N. Spergel, and G.D. Starkman, Class. Quantum Grav.
\textbf{15}, 2657 (1998).

\bibitem{Ford80a} L.H. Ford, Phys. Rev. D \textbf{22}, 3003 (1980).

\bibitem{Ford79} L.H. Ford and T. Yoshimura, Phys. Lett. A \textbf{70}, 89
(1979).

\bibitem{Toms80a} D.J. Toms, Phys. Rev. D \textbf{21}, 928 (1980).

\bibitem{Toms80b} D.J. Toms, Phys. Rev. D \textbf{21}, 2805 (1980).

\bibitem{Odin88} S.D. Odintsov, Sov. J. Nucl. Phys. \textbf{48}, 1148
(1988); I.L. Buchbinder and S.D. Odintsov, Int. J. Mod. Phys. \textbf{A4},
4337 (1989); Fortschr. Phys. \textbf{37}, 225 (1989); I.L.Buchbinder, S.D.
Odintsov, and V.P. Dergaleo, Theor. Math. Phys. \textbf{80}, 150 (1989).

\bibitem{Most97} V.M. Mostepanenko and N.N. Trunov, \textit{The Casimir
Effect and Its Applications} (Clarendon, Oxford, 1997).

\bibitem{Bord01} M. Bordag, U. Mohidden, and V.M. Mostepanenko, Phys. Rep.
\textbf{353}, 1 (2001); K.A. Milton, \textit{The Casimir Effect: Physical
Manifestation of Zero-Point Energy} (World Scientific, Singapore, 2002); E.
Elizalde, S.D. Odintsov, A. Romeo, A.A. Bytsenko and S. Zerbini, \textit{%
Zeta regularization techniques with applications} (World Scientific,
Singapore, 1994); M.J. Duff, B.E.W. Nilsson, and C.N. Pope, Phys. Rep.
\textbf{130}, 1 (1986); A.A. Bytsenko, G. Cognola, L. Vanzo, and S. Zerbini,
Phys. Rep. \textbf{266}, 1 (1996).

\bibitem{Milt03} K.A. Milton, Grav. Cosmol. \textbf{9}, 66 (2003).

\bibitem{Eliz06} E. Elizalde, J. Phys. A \textbf{39}, 6299 (2006).

\bibitem{Gree07} B. Green and J. Levin, JHEP \textbf{11}, 096 (2007); P.
Burikham, A. Chatrabhuti, P. Patcharamaneepakorn, and K. Pimsamarn,
arXiv:0802.3564.

\bibitem{Lind90} A.D. Linde, \textit{Particle Physics and Inflationary
Cosmology} (Harwood Academic Publishers, Chur, Switzerland 1990).

\bibitem{Ries07} A.G. Riess et al., Astrophys. J. \textbf{659}, 98 (2007);
D.N. Spergel et al., Astrophys. J. Suppl. Ser. \textbf{170}, 377 (2007);
U.~Seljak, A. Slosar, and P. McDonald, JCAP \textbf{0610}, 014 (2006); E.
Komatsu et al., arXiv:0803.0547.

\bibitem{Kach03} S. Kachru, R. Kallosh, A. Linde, and S.P. Trivedi, Phys.
Rev. D \textbf{68}, 046005 (2003).

\bibitem{Silv07} E. Silverstein, arXiv:0712.1196.

\bibitem{McIn04} B. McInnes, Nucl. Phys. B \textbf{709}, 213 (2005); B.
McInnes, Nucl. Phys. B \textbf{748} , 309 (2006).

\bibitem{Zeld84} Y.B. Zeldovich and A.A. Starobinsky, Sov. Astron. Lett.
\textbf{10}, 135 (1984).

\bibitem{Gonc85} Yu.P. Goncharov and A.A. Bytsenko, Phys. Lett. B \textbf{160%
}, 385 (1985); Yu.P. Goncharov and A.A. Bytsenko, Phys. Lett. B \textbf{169}%
, 171 (1986); Yu.P. Goncharov and A.A. Bytsenko, Nucl. Phys. B \textbf{271},
726 (1986); Yu.P. Goncharov and A.A. Bytsenko, Class. Quant. Grav. \textbf{4}%
, 555 (1987).

\bibitem{Saha07} A.A. Saharian and M.R. Setare, Phys. Lett. B \textbf{659},
367 (2008).

\bibitem{dSbrane} S. Nojiri, S. D. Odintsov, and S. Zerbini, Class. Quantum.
Grav. \textbf{17}, 4855 (2000); W. Naylor and M. Sasaki, Phys. Lett. B
\textbf{542}, 289 (2002); E. Elizalde, S. Nojiri, S. D. Odintsov, and S.
Ogushi, Phys. Rev. D \textbf{67}, 063515 (2003); I.G. Moss, W. Naylor, W.
Santiago-German, and M. Sasaki, Phys. Rev. D \textbf{67}, 125010 (2003); K.
Uzawa, Prog. Theor. Phys. \textbf{110}, 457 (2003); A. A. Saharian and M. R.
Setare, Phys. Lett. B \textbf{584}, 306 (2004); M.R. Setare, Phys. Lett. B
\textbf{620}, 111 (2005); M. Minamitsuji, W. Naylor, and M. Sasaki, Nucl.
Phys. B \textbf{737}, 121 (2006); M.R. Setare, Phys. Lett. B \textbf{637}, 1
(2006).

\bibitem{Flac03} A. Flachi, J. Garriga, O. Pujol\`{a}s, and T. Tanaka, J.
High Energy Phys. \textbf{08}, 053 (2003); A. Flachi and O. Pujol\`{a}s,
Phys. Rev. D \textbf{68}, 025023 (2003); A.A. Saharian, Phys. Rev. D \textbf{%
73}, 044012 (2006); A.A. Saharian, Phys. Rev. D \textbf{73}, 064019 (2006);
A.A. Saharian, Phys. Rev. D \textbf{74}, 124009 (2006).

\bibitem{Bell08} S. Bellucci, A.A. Saharian, arXiv:0802.2190.

\bibitem{Cand75} P. Candelas and D.J. Raine, Phys. Rev. D \textbf{12}, 965
(1975).

\bibitem{Dowk76} J.S. Dowker and R. Critchley, Phys. Rev. D \textbf{13}, 224
(1976); J.S. Dowker and R. Critchley, Phys. Rev. D \textbf{13}, 3224 (1976).

\bibitem{Bunc78} T.S. Bunch and P.C.W. Davies, Proc. R. Soc. London \textbf{%
A360}, 117 (1978).

\bibitem{Birr82} N.D. Birrell and P.C.W. Davies, \textit{Quantum Fields in
Curved Space} (Cambridge University Press, Cambridge, 1982).

\bibitem{Vile82} A. Vilenkin and L.H. Ford, Phys. Rev. D \textbf{26}, 1231
(1982).

\bibitem{Alle83} B. Allen, Nucl. Phys. B \textbf{226}, 228 (1983).

\bibitem{Birr79} N.D. Birrell, J. Phys. A \textbf{12}, 337 (1979).

\bibitem{Isha78} C.J. Isham, Proc. R. Soc. London \textbf{A362}, 383 (1978);
C.J. Isham, Proc. R. Soc. London \textbf{A364}, 591 (1978).

\bibitem{Bere82} V.B. Berestetskii, E.M. Lifshits, and L.P. Pitaevskii,
\textit{Quantum Electrodynamics} (Pergamon, Oxford, 1982).

\bibitem{Cota02} I.I. Cot\u{a}escu, Phys. Rev. D \textbf{65}, 084008 (2002).

\bibitem{Saha07Gen} A. A. Saharian, "The Generalized Abel-Plana Formula with
Applications to Bessel Functions and Casimir Effect," Report No.
IC/2007/082; arxiv:0708.1187.

\bibitem{Prud86} A. P. Prudnikov, Yu.A. Brychkov, and O. I. Marichev,
\textit{Integrals and Series, Vol. 2: Special Functions} (Gordon and Breach,
New York, 1986).
\end{thebibliography}
\end{document}